\documentclass[12pt]{iopart}
\usepackage{graphicx}
\begin{document}

\title[What do we learn from Resonance Production in Heavy Ion Collisions?]
{What do we learn from Resonance Production in Heavy Ion
Collisions?}

\author{Christina Markert
\footnote[3]{(Christina.markert@yale.edu)} }

\address{Physics Department, Yale University, New Haven, CT 06520,
USA}


\begin{abstract}
Resonances with their short life time and strong coupling to the
dense and hot medium are suggested as a signature of the early
stage of the fireball created in a heavy ion collision
\cite{rap00,lut01,lut02}. The comparison of resonances with
different lifetimes and quark contents may give information about
time evolution and density and temperature of during the expanding
of fireball medium. Resonances in elementary reactions have been
measured since 1960. Resonance production in elementary collisions
compared with heavy ion collisions where we expect to create a hot
and dense medium may show the direct of influence of the medium on
the resonances. This paper shows a selection of the recent
resonance measurements from SPS and RHIC heavy ion colliders.
\end{abstract}


\section{Resonances in Medium}

In a heavy ion collision an extended hot and dense fireball medium
is created. During the expansion of the fireball two freeze-out
conditions are defined, chemical and thermal, representing the end
of the inelastic and elastic interactions. In a dynamical evolving
system produced resonances decay and may get regenerated. Decay
daughters of resonances which decay inside the medium may also
scatter with other particles from the medium, mostly pions for SPS
and RHIC energies. This results in a non-reconstructable resonance
from the decay daughters measured in the detector, because the
invariant mass of the decay daughters no longer matches that of
the parent. The reconstructed resonance signal from the scattered
decay daughters is a few hundred MeV broad in the UrQMD model
\cite{ble02,ble02b} and is therefor not distinguishable from the
background distribution. The rescattering and regeneration
(pseudo-elastic) process for resonances and their decay particles
depend on the individual cross sections and are dominant after
chemical but before the kinetic freeze-out. These interactions can
result in changes of the reconstructed resonance yields, momentum
spectra and widths. Rescattering will decrease the measured
resonance yields while the mechanism of regeneration will increase
the them. Microscopic model calculations include every step in a
heavy ion interaction in terms of elastic and inelastic
interactions of hadrons and strings \cite{ble02,ble02b}. They are
therefore able to describe the rescattering and regeneration
contribution on the resonances in a fireball. The prediction of
this model (UrQMD) is a signal loss for some of the resonances due
to more rescattering than regeneration in the low momentum region
p$_{\rm T}<1$~GeV for the hadronic decays. The leptonic decay
products are not significantly affected by rescattering due to
their low cross section with hadrons. Comparisons between the
yield and momentum spectra of the hadronic and leptonic decay
channels can indicate the magnitude of the rescattering and
regeneration contribution between chemical and thermal freeze-out.

\section{Resonances in elementary collisions}

\begin{figure}[h]
\centering
 \vspace{0.5cm}
\includegraphics[width=0.45\textwidth]{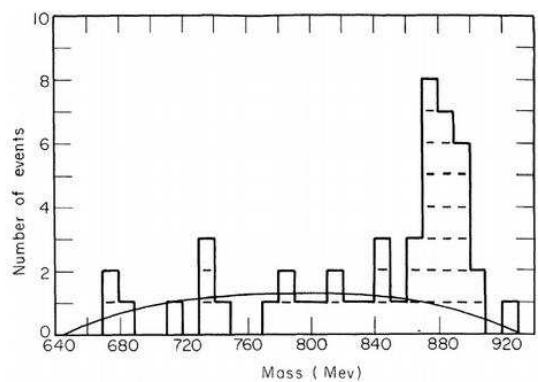}
 \caption{First discovery of resonance in invariant mass reconstruction
 spectrum of K(892)$^{-}$ in K$^{0}$ and pion
 channel \cite{als61}.}
\label{discovery}
\end{figure}

A resonance is a particle with a higher mass than the
corresponding ground state particle with the same quark content.
It decays strongly and therefore the lifetime, $\tau$, of the
resonance is short (within a few fm/c). This causes a wide spread
in energy and a natural width, given by $\Gamma$ = $\hbar$/$\tau$.
These broad states with finite $\Gamma$ and lifetime $\tau$ can be
formed by collisions between the same particles into which they
decay. The resonances can have hadronic and leptonic decay
channels. Since they have a short lifetime they can only be
measured by reconstruction using the decay particles measured in
the detector. \newline
 At the beginning of the 1960's the first
resonance particle was discovered in a bubble chamber experiment
at Berkeley using a kaon beam hitting a proton target. The
K(892)$^{-}$ resonance is formed by a K$^{-}$+p $\rightarrow$
K(892)$^{-}$+p reaction \cite{als61}. The reconstruction of the
K(892)$^{-}$ was done via invariant mass calculation using the
decay products of a K(892)$^{-}$~$\rightarrow$ K$^{0}$ + $\pi^{-}$
decay (see Fig~\ref{discovery}). The signal in the invariant mass
spectrum peaks around the mass of 890 MeV/c$^{2}$ for the
K$^{-}$(892) particle. In 1968 Luis Walter Alvarez received the
Nobel Prize for resonance particles discovered in 1960.

\begin{figure}[h]
\centering
 \vspace{0.5cm}
\includegraphics[width=0.60\textwidth]{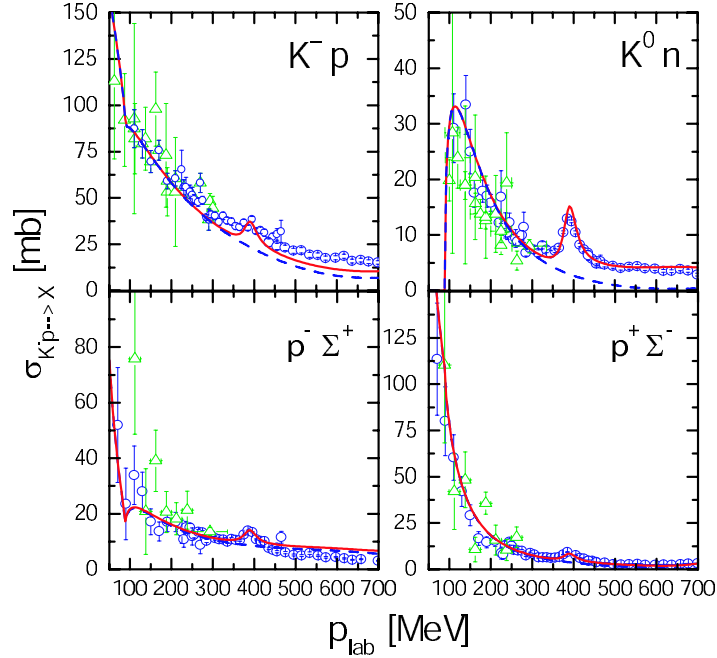}
 \caption{First discovery of resonance in invariant mass reconstruction
 spectrum of K(892)$^{-}$ in K$^{0}$ and pion
 channel \cite{mas76}.}
\label{cross}
\end{figure}

Later experiments observed more resonances, and using an energy
scan of the kaon beam measured the total (elastic and inelastic)
cross section of the K$^{-}$+p scattering. The scattering
amplitudes versus the kaon beam momentum for different decay
channels of the $\Lambda$(1520) resonance are shown in
Fig.~\ref{cross}. At the kaon beam momentum of 395 MeV/c the cross
section peaks in the kinematic region of the $\Lambda$(1520)
resonance. Theoretical in-medium calculations made using the
relativistic chiral SU(3) Lagrangian to describe the resonances by
using hadrons instead of quarks and gluons are in good agreement
with the rescattering amplitude in Fig.~\ref{cross} \cite{lut02}.
The solid line is the contribution of the s-, p-, d-waves
functions while the dashed line shows only the contribution from
the s-wave. The mass and width predictions for resonances within
this model will be discussed later in this paper.

\section{Resonance Reconstruction }

The resonances are reconstructed from their observed decay
daughters. The $\Lambda$s from a $\Sigma$(1385) and $\Xi$(1530)
decay are reconstructed via topological analysis (see
table~\ref{resotable}). The resonance signal is obtained by the
invariant mass reconstruction of each daughter combination and
subtraction of the combinatorial background calculated by mixed
event or like-signed techniques. The resonance ratios, spectra and
yields are measured at mid-rapidity for RHIC at $\sqrt{s_{\rm NN}}
= $ 200 GeV and over 4$\pi$ for SPS at $\sqrt{s_{\rm NN}} = $ 17
GeV. The central trigger selection for Au+Au collisions at RHIC
takes the 5\% or 10\% and for Pb+Pb collisions at SPS the 5\% of
the most central inelastic interactions. The setup for the
proton+proton interaction is a minimum bias trigger.

\begin{table}[htb]
\begin{center}
\begin{tabular}{lllll}
\hline
Particle & mass [MeV/ c$^{2}$]& width [MeV/ c$^{2}$]& lifetime [fm/c]& decay channel \\
\hline

$\Delta$(1232) & 1232 $\pm$ 2 & 120 $\pm$ 5& 1.6 & p + $\pi$ \\
K(892) & 896.1 $\pm$ 0.27 &  50.7 $\pm$ 0.6 & 3.89 & K + $\pi$ \\
$\Sigma$(1385) & 1385 $\pm$ 3 &  37 $\pm$ 2 & 5.2 & $\Lambda$($\rightarrow$p+$\pi$) + $\pi$ \\
$\Lambda$(1520)   & 1519.5 $\pm$ 1.0 &  15.6 $\pm$ 1.0 & 12.6 & p + K \\
$\phi$(1020) & 1019.417 $\pm$ 0.014 &  4.458 $\pm$ 0.032 &  44.6 & $K^{+}$  + $K^{-}$ \\
 \hline
\end{tabular}
\end{center}
\vspace{-0.4cm}
 \caption[]{Resonances and their main hadronic decay channels from PDG \cite{pdg98}}
\label{tab1} \label{resotable}
\end{table}

\section{Rescattering and Regeneration}

Rescattering will decrease the resonance signal in the invariant
mass reconstruction while the regeneration process will increase
the signal. If the signal loss or gain is due to rescattering or
regeneration a comparison of the resonance yields and momentum
spectra with thermal models is not appropriate as the contribution
of the rescattering phase has taken to be into account.
Microscopic models (such as UrQMD) are able to describe such a
phase \cite{ble02,ble02b}. Depending on the lifetime and the
rescattering and regeneration cross sections the observed
resonances come from different times of the fireball source.

\begin{figure}[h]
\centering
 \vspace{0.5cm}
\includegraphics[width=1.0\textwidth]{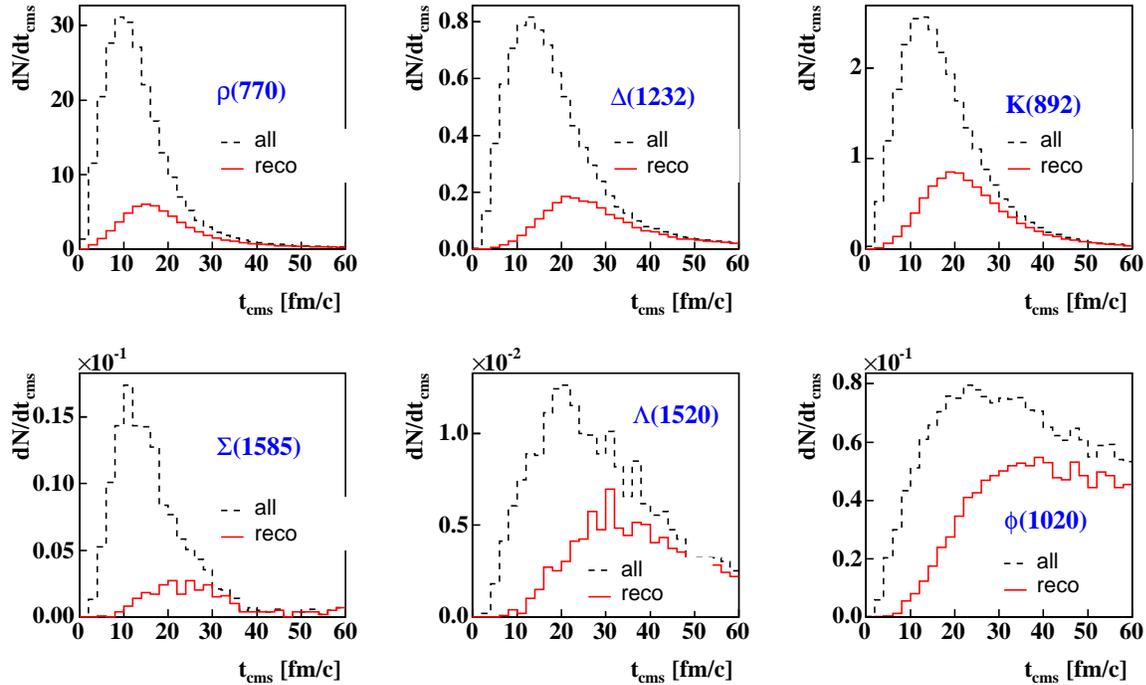}
 \caption{In UrQMD produced (dashed line) and reconstrucable (solid line) resonances versus the fireball time for
$\rho$(770), $\Delta$(1232), K(892), $\Sigma$(1385),
$\Lambda$(1520) and $\phi$(1020) at mid-rapidity for Au+Au
collisions at $\sqrt{s_{\rm NN}}$ =  200 GeV.} \label{urqmdsurv}
\end{figure}

\begin{figure}[h]
\centering
\begin{minipage}[b]{0.32\linewidth}
\includegraphics[width=1.1\textwidth]{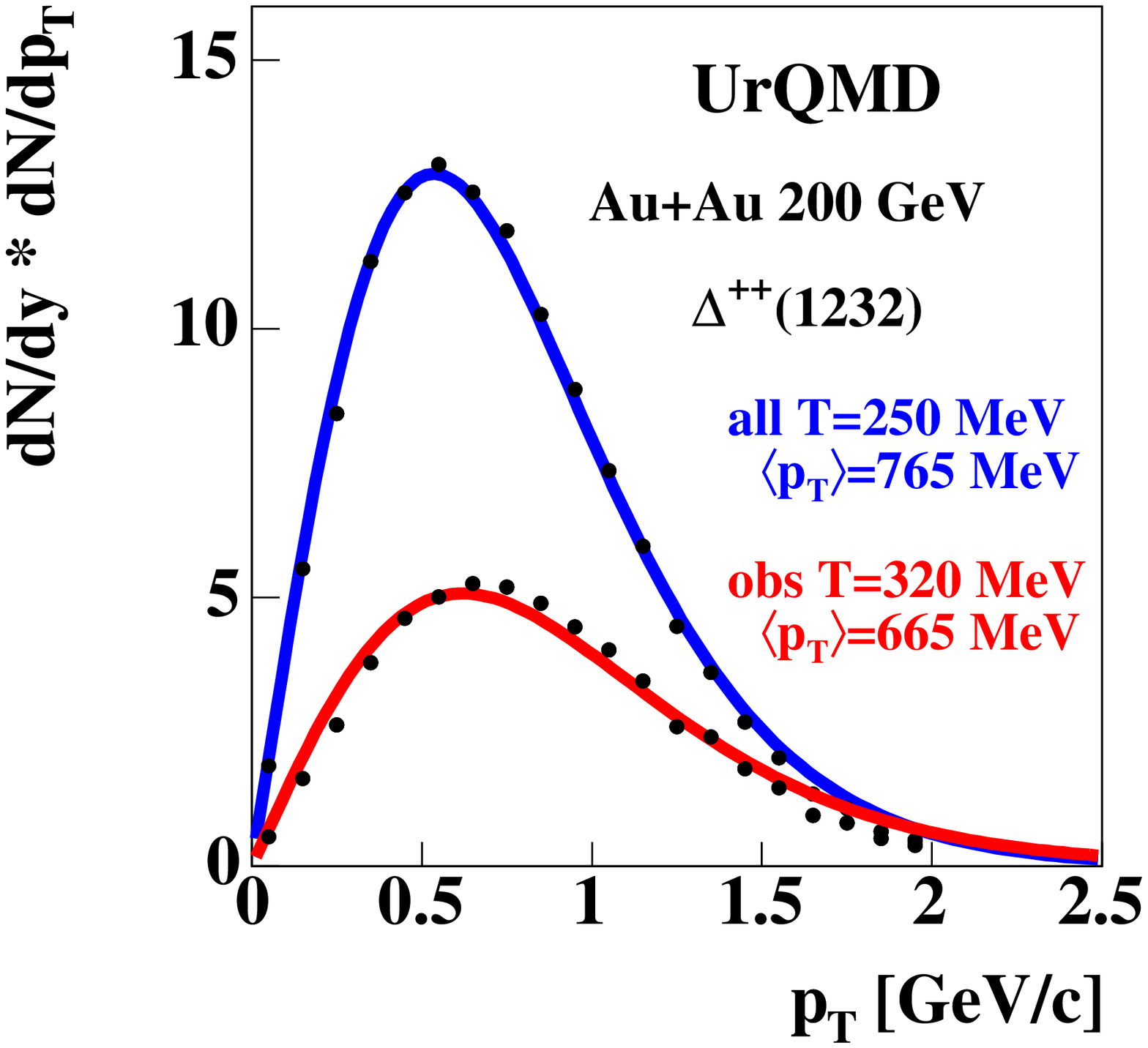}
\end{minipage}
\begin{minipage}[b]{0.32\linewidth}
\includegraphics[width=1.1\textwidth]{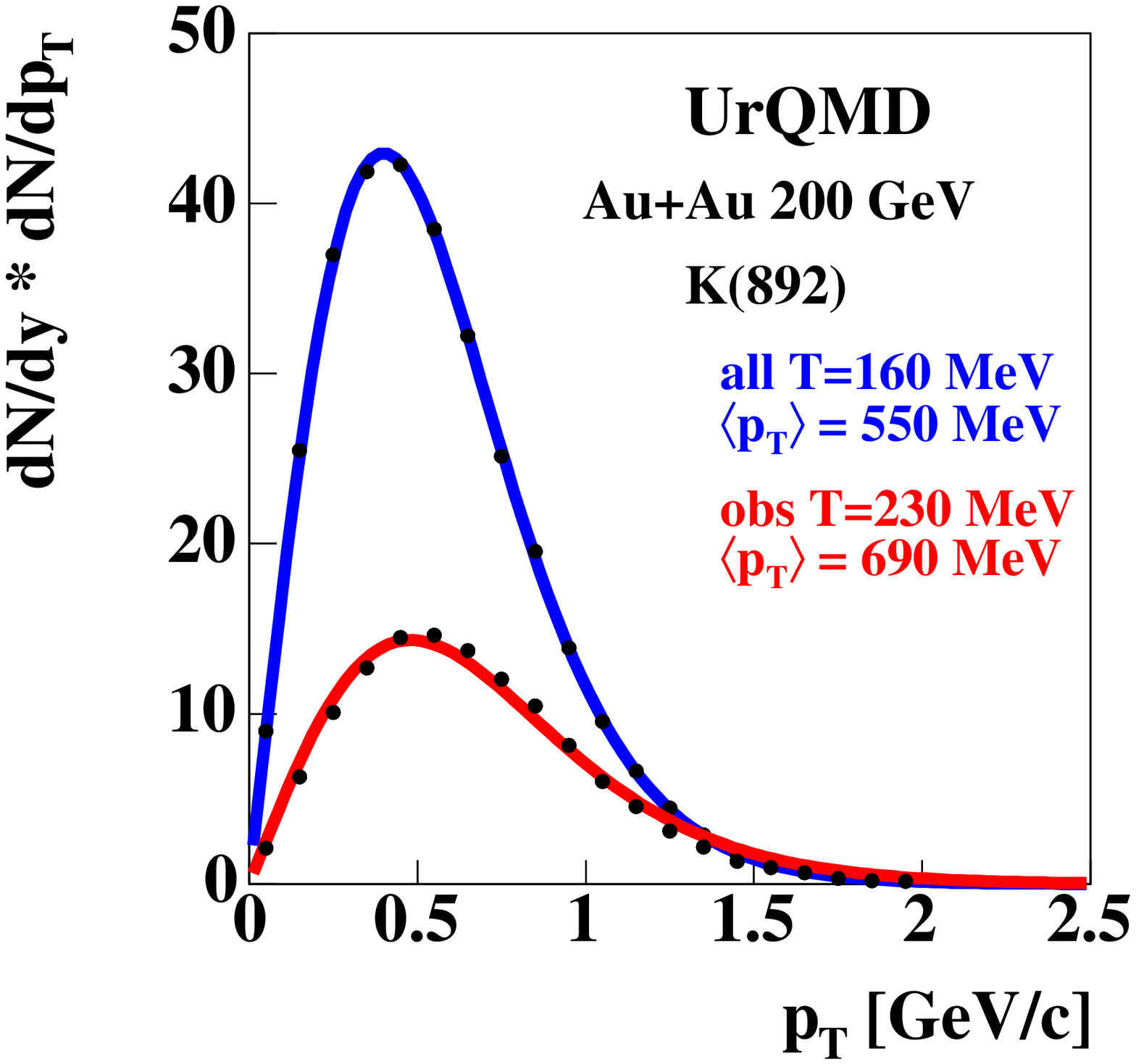}
\end{minipage}
\begin{minipage}[b]{0.32\linewidth}
\includegraphics[width=1.1\textwidth]{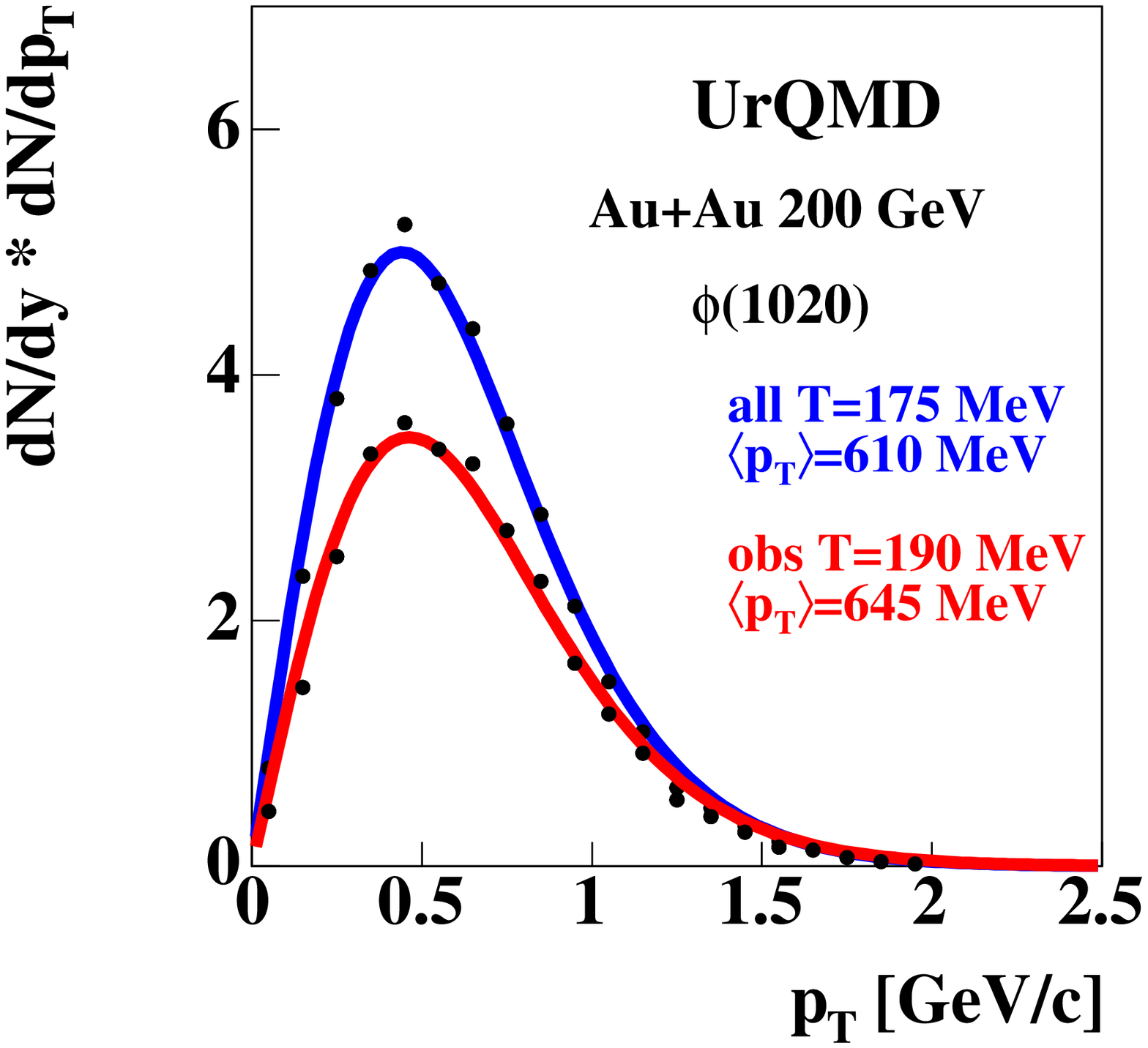}
\end{minipage}
 \caption{Transverse momentum spectrum of produced and reconstrucable resonances in UrQMD
 $\Delta$(1232), K(892) and $\phi$(1020) at mid-rapidity for Au+Au collisions at $\sqrt{s_{\rm NN}}$~=~200 GeV.}
 \label{urqmdsurvmt}
\end{figure}

Fig.~\ref{urqmdsurv} shows the UrQMD produced (dashed line) and
reconstructable (solid line) resonances versus the fireball time
for different resonances with different lifetimes: $\tau_{\rho}$=
1.3 fm/c, $\tau_{\Delta}(1232)$ = 1.7 fm/c, $\tau_{K(892)}$= 4.0
fm/c, $\tau_{\Sigma(1385)}$= 5.5 fm/c, $\tau_{\Lambda(1520)}$ = 13
fm/c and $\tau_{\phi}$ = 46 fm/c. We observe two features: the
mean of the fireball lifetime at the decay of the reconstructable
resonances scales with the lifetime of the resonance and observed
resonances compared to those produced is higher for states with
less rescattering and regeneration. Note that this ratio is not
directly applicable to the total yield of produced resonances from
a thermal model prediction because it is a dynamical process where
resonances decay and get generated several times during the whole
expansion time which makes this ratio very small. However, the
absolute values of the measurable resonances of UrQMD can be
compared with the experimental results. One short remark here is
that UrQMD has a long lifetime for a Au+Au heavy ion reaction
(longer then 30 fm/c). It is not clear how the resonance
production in terms of rescattering and regeneration would be
affected if the source would expand faster. \\
From UrQMD calculations we learn that the signal loss of
resonances due to rescattering takes place predominantly in the
low momentum region \cite{ble02,ble02b}. Therefore the
reconstructed transverse momentum spectra are expected to change
to higher inverse slope parameters and higher $\langle$p$_{\rm
T}$$\rangle$. An exponential fit to the transverse momentum of the
created and measurable resonances calculated in UrQMD shows that
the spectra shape looks thermal before and after rescattering and
regeneration of the resonance Fig~\ref{urqmdsurvmt}. The increase
in $\langle$p$_{\rm T}$$\rangle$ is for the $\Delta$(1232) and
K(892) is of the order of 100 MeV/c and for the $\phi$(1020) 35
MeV/c.\\

\begin{figure}[h]
\centering
 \vspace{0.5cm}
\includegraphics[width=0.50\textwidth]{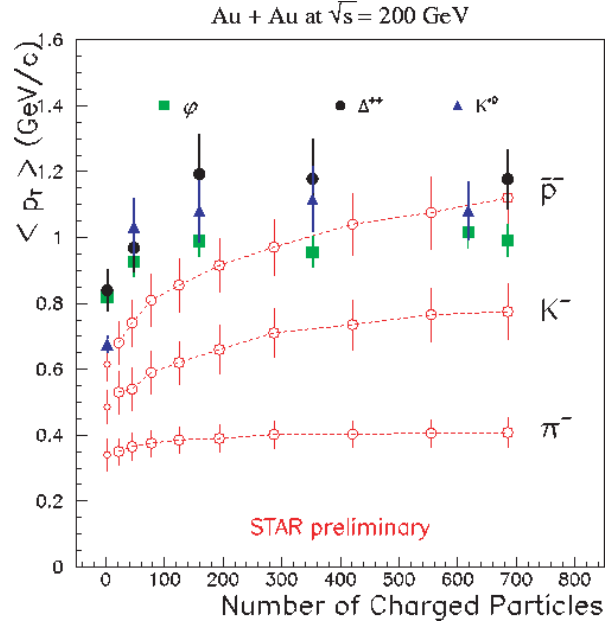}
 \caption{ The $\langle$p$_{\rm T}$$\rangle$ for resonances
 and ground state particles in p+p and Au+Au collisions
 versus number of charged particles \cite{ma04,mar04,zha04,sal04a}.}
 \label{resopt}
\end{figure}

We also would expect a higher increase of inverse slope and
$\langle$p$_{\rm T}$$\rangle$ for resonances from p+p to Au+Au
collisions than for ground state particles due to the rescattering
and regeneration effect in the surrounding fireball medium even in
peripheral collisions. The STAR data from p+p and Au+Au collisions
at $\sqrt{s_{\rm NN}} = $ 200 GeV confirm this trend of a strong
increase $\langle$p$_{\rm T}$$\rangle$ for resonances from p+p to
peripheral Au+Au collisions which is not present for the ground
state particles (see Fig~\ref{resopt}) \cite{ma04,mar04,zha04}.
Due to this observation one would also expect a stronger deviation
of the transverse momentum spectra of the resonances compared to
ground state particles in the low momentum region of a thermal
model.

\section{Resonance Yields}

Under the assumption that all the particles freeze out at the same
time their yields are extracted from a thermal model for one fit
parameter set of chemical temperature T$_{ch}$, chemical
potentials $\mu_{b}$ and $\mu_{s}$ and the strangeness saturation
factor $\gamma_{s}$. This chemical freeze-out is the end of the
inelastic interactions where the yields of most particle species
are constant. Elastic and pseudo-elastic interactions do not
change the yield of the ground state particles. The momentum
spectra can change due to further elastic interactions.
Resonances, which are measured (reconstructed) by their decay
daughters, are affected by the pseudo-elastic interactions which
can cause a change in the yield of the measured signal in the
invariant mass spectrum and a change in the momentum distribution
of the reconstructed resonance. The yields and momentum spectra
will change for resonances according to their cross section for
rescattering and regeneration of resonances and rescattering of
their decay daughters, the resonance lifetime and the medium
density. Regeneration may compensate the effects of rescattering.
Yields and momentum spectra in comparison with the prediction from
a thermal model may suggest the magnitude of influence of
rescattering over regeneration. Resonance over non-resonance
particle ratios of heavy ion collisions compared to p+p
interactions may indicate signal yield changes due to rescattering
and regeneration processes. Table~\ref{statmodel} shows measured
particle ratios in p+p and Au+Au collisions at $\sqrt{s_{\rm NN}}
= $ 200 GeV and the expected particle ratios for the resonances
from a thermal model based on a fit to the ground state particle
ratios for Au+Au collisions \cite{pbm01}.

 \vspace{0.7cm}

\begin{table}[htb]
\begin{center}
\begin{tabular}{l||l||l|l}
\hline
Particle ratio & p+p data & Au+Au data & Au+Au model \\
\hline
$\Delta$(1232)/p & 0.72 $\pm$ 0.108  &   0.96 $\pm$ 0.148  &  0.68 \\
K(892)$^{0}$/K & 0.389  $\pm$  0.029  &  0.228  $\pm$ 0.044  &  0.32\\
$\Lambda$(1520)/$\Lambda$ & 0.104 $\pm$ 0.03 &   0.033 $\pm$ 0.017   & 0.071\\
$\phi$(1020)/K & 0.124 $\pm$ 0.025   & 0.1556 $\pm$ 0.0311  &  0.13  \\
 \hline
\end{tabular}
\end{center}
 \caption[]{Comparison of STAR data with thermal model predictions of resonance/non-resonance ratios in p+p
and Au+Au collisions at $\sqrt{s_{\rm NN}} = $ 200 GeV
\cite{pbm01}.}
 \label{statmodel}
\end{table}
 \vspace{0.7cm}

The K(892)/K and the $\Lambda$(1520)/$\Lambda$ in Au+Au collisions
are lower than in p+p collisions and the expected values from the
thermal model are also higher than the data ( see Fig~\ref{part}).
These data suggest that the rescattering cross section is larger
than the regeneration from the K+$\pi$ and K+p. The K(892)
lifetime is smaller than the lifetime of the $\Lambda$(1520) which
would imply a larger suppression for K(892)/K than for the
$\Lambda$(1520)/$\Lambda$ ratio. The competing contribution of
regeneration seems to be larger for K(892) than for the
$\Lambda$(1520). The $\phi$(1020)/K ratio is consistent within
errors with the thermal model prediction, which is expected
because only a small fraction of the $\phi$(1020) are decaying
inside the fireball due to the long lifetime of the $\phi$(1020),
of 46 fm/c. The expected contribution of rescattering for the
short lived $\Delta$(1232) resonance is larger than that for the
K(892) and the $\Lambda$(1520). However the $\Delta$(1232)/p ratio
does not decrease from p+p to Au+Au collisions and is on the order
of 41\% $\pm$ 22\% higher than the thermal model prediction. This
indicates a large cross section for the regeneration of
$\Delta$(1232) resonance in the p+$\pi$ channel. The first
$\Sigma$(1385) yields from heavy ion collisions appear to follow
the same trend as the $\Delta$(1232) \cite{sal04}. This implies
that the $\Lambda$+$\pi$ regeneration cross section is
compensating the signal loss from rescattering. From this
observation we can conclude that there is a ranking order of the
cross section for the different regeneration processes:
$\sigma_{p+\pi}$ $\geq$ $\sigma_{\Lambda+\pi}$ $>$
$\sigma_{K+\pi}$ $>$ $\sigma_{K+p}$. The first attempt to measure
the $\Xi(1530)$ with the STAR detector are shown in [14]. Final
results will give contributions to the $\Xi(1530)$ + $\pi$ cross
section and add a more stringent test to theoretical descriptions
of the data.

\section{Time Scale}

\begin{figure}[htb]
\centering
 \vspace{0.5cm}
\includegraphics[width=0.70\textwidth]{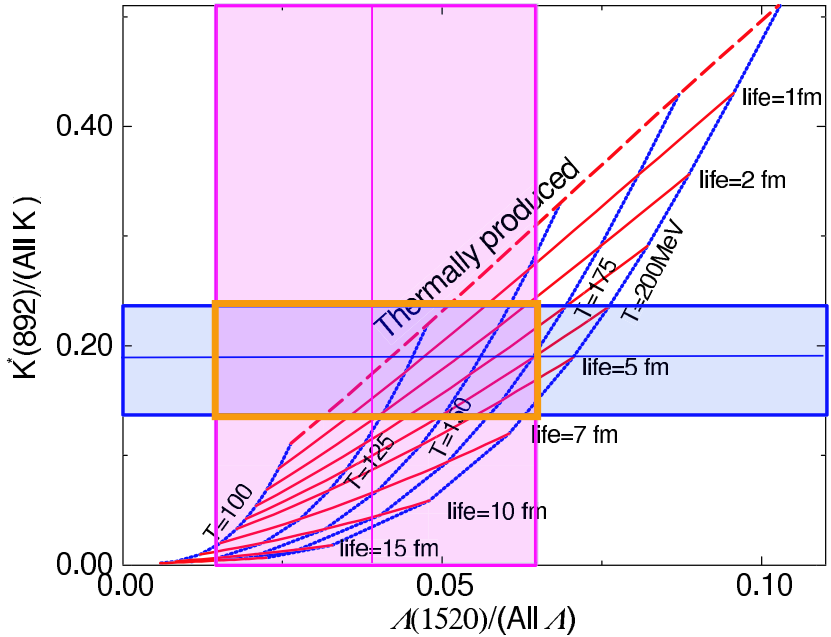}
 \caption{Dependence of lifetime and chemical freeze out temperature of a fireball
 source given by the two particle ratios $\Lambda$(1520)/$\Lambda$ and
K(892)/K including limits from the STAR data for cental Au+Au
collisions at $\sqrt{s_{\rm NN}} = $ 200 GeV \cite{tor01,tor01a}.}
\label{rafelski}
\end{figure}

Depending on the length of the time interval between chemical and
kinetic freeze-out, $\Delta \tau$ , the magnitude of the
suppression factor of the measured resonance will change due to
contributions from rescattering and regeneration. A model using
thermally produced particle yields at chemical freeze-out and an
additional rescattering phase, including the lifetime of the
resonances and decay product interactions within the expanding
fireball of nuclear matter, can place a lower limit on
\cite{tor01,tor01a,mar02}. The probability of rescattering is
described by the (energy averaged) interaction cross section of
the decay particles of the resonances within the medium, and
depends on the radius, density and velocity of the fireball. A
slower expansion of the fireball would lead to a higher
suppression of the resonance signal due to the larger rate of
rescattering. This model does not include regeneration and is
therefore predicting a lower limit of the lifetime $\Delta \tau$
between the two freeze-out surfaces. It is only applicable if the
contribution from rescattering is larger than from regeneration.
The two ratios K(892)/K and $\Lambda$(1520)/$\Lambda$ are expected
to have a larger rescattering contribution. A $\Delta\tau$ $>$
4~fm/c results if chemical freeze-out occurs at 160 MeV (See
Fig~\ref{rafelski}.

\begin{figure}[htb]
\centering
\includegraphics[width=0.63\textwidth]{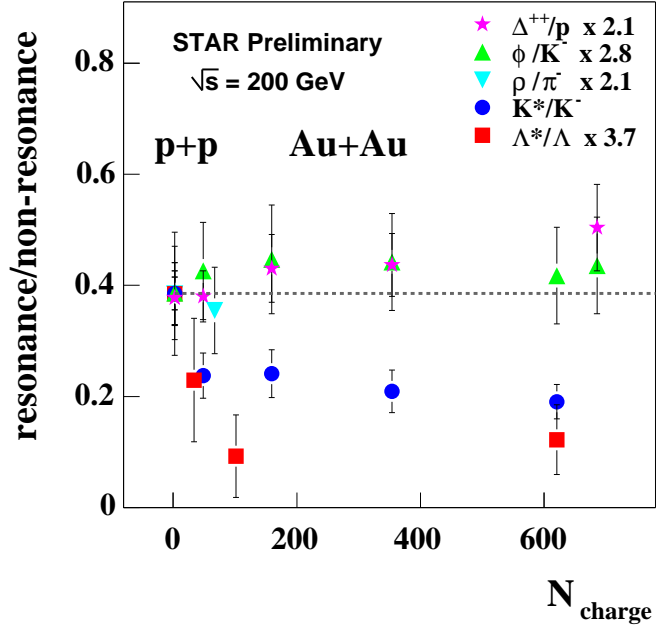}
\caption{Resonance/non-resonance ratios of $\phi$(1020)/K$^{-}$
\cite{ma04}, $\Delta(1232)^{++}$/p, $\rho(770)/\pi$ \cite{fac04},
K(892)/K$^{-}$ \cite{zha04} and $\Lambda$(1520)/$\Lambda$)
\cite{gau04,mar03} for p+p and Au+Au collisions at $\sqrt{s_{\rm
NN}}$ =  200 GeV. The ratios are normalized to the K(892)/K$^{-}$
p+p ratio. Statistical and systematical errors are included.}
 \label{part}
\end{figure}

This model gives every combination between temperature and life
span in the marked area of the data and their errors. Another
valid interpretation of the data in this model is a chemical
freeze-out temperature of T=110~$\pm$ 10 MeV and a time span of
$\Delta\tau$ $=$ 0~fm/c. With smaller errors the combination of T
and $\Delta \tau$ can be determined more precisely.
Fig.~\ref{part} shows the resonance/non-resonance ratios in p+p
and Au+Au collisions for different centralities. The suppression
of the K(892)/K and the $\Lambda$(1520)/$\Lambda$ in Au+Au
collisions sets in at the peripheral collisions and remains
constant up to the central collisions. This would suggest the same
life span between chemical and thermal freeze-out in
mid-peripheral and central Au+Au collisions.

\section{Leptonic and Hadronic Decay Channels}

\begin{figure}[htb]
\vspace{0.8cm}
 \centering
\includegraphics[width=0.6\textwidth]{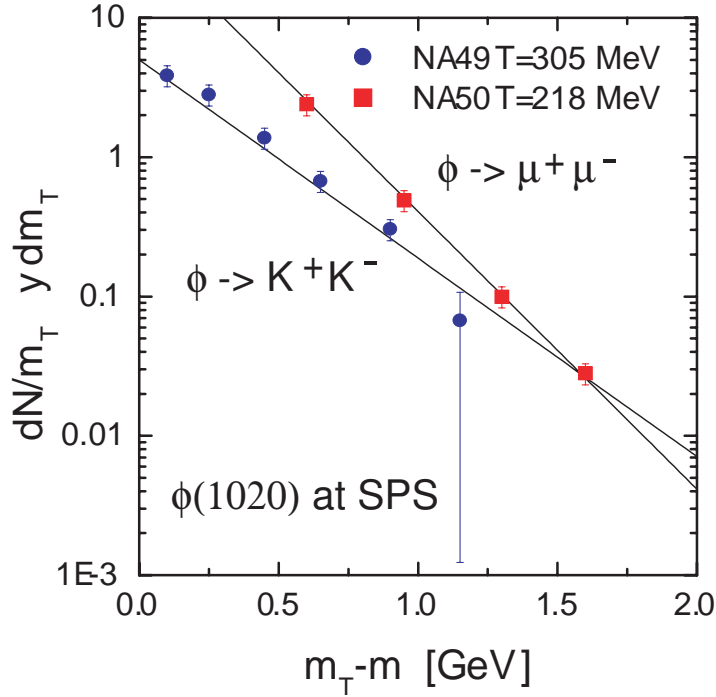}
\caption{Transverse momentum distribution of the hadronic decay
$\phi$(1020) $\rightarrow$ K$^{+}$ + K$^{-}$ from NA49
\cite{fri97} and the leptonic decay $\phi$(1020) $\rightarrow$
$\mu^{+}$ + $\mu^{-}$ from NA50 \cite{wil99}.}
 \label{phi}
\end{figure}

Direct comparisons of the spectra and yields in heavy ion
collisions from leptonic and hadronic decay channels may show the
influence of the hadronic interaction phase after chemical
freeze-out folded with the initial production of the $\phi$(1020)
at the early stage. The $\phi$(1020) is one of the resonances
where we have measurements from the leptonic and hadronic decay
channels in heavy ion collisions. At the moment there are only SPS
data from heavy ion collisions available for the leptonic channel
of $\phi$(1020). Results from $\phi$(1020) production in Au+Au
collisions at RHIC energies into e$^{+}$ + e$^{-}$ should be
measured in the next months. From SPS we have the so-called
'$\phi$ puzzle' with two different momentum distributions and
yields for the leptonic and hadronic decay. Fig~\ref{phi} shows
the transverse momentum distribution of the hadronic decay $\phi$
$\rightarrow$ K$^{+}$ + K$^{-}$ from NA49 and the leptonic decay
$\phi$ $\rightarrow$ $\mu^{+}$ + $\mu^{-}$ from NA50
\cite{fri97,wil99}. The inverse slope parameter from fits to the
momentum spectra indicated as lines are T~=~305~$\pm$~15~MeV for
hadronic decay and T~=~218~$\pm$~10~MeV for leptonic decay. The
extracted yield from the momentum spectrum of the leptonic decay
is a factor of 4~$\pm$~2 higher than the one for the hadronic
decay. Microscopic calculations (UrQMD) estimated for the
$\phi$(1020) resonance a suppression of 20-30\% of the yield in
the hadronic decay channel due to rescattering of the kaon decay
daughters in the low momentum region p$_{\rm T}$~$<$~1~GeV
\cite{ble02,ble02b}. The rescattering is negligible for the
leptonic decay due to the very low cross section of interaction
with the hadronic phase. The trend of the lower signal in the low
momentum region of the hadronic decay compared to the leptonic
decay from the data is in agreement with the model. However this
signal loss of 20-30\% from the model calculation is not
sufficient to explain the factor of 4~$\pm$~2 in the measured
yield of the data.


This gives room for possible medium effects on the resonance
production which is effective at an earlier stage, before chemical
freeze-out. An approach to describe the in medium modification of
the $\phi$(1020) resonance has been done by K. Haglin and E.
Kolomeitsev \cite{hag04,kol99}. This attempt to fit the SPS data
describes a hot and dense fireball where the lifetime of the
$\phi$(1020) resonance is modified towards smaller lifetimes and
therefore more of the $\phi$(1020) resonances decay inside the
medium. This will introduce a larger signal loss due to
rescattering of the hadronic decay daughters. In this conference
the $\phi$(1020) measurements from PHENIX in d+Au \cite{pal04} and
in Au+Au \cite{muk04} and in d+Au collision systems were shown.

\section{Width and Mass}

Theoretical in-medium calculations of resonances predict mass and
width changes depending on the nuclear matter density. The theory
of Hendrik van Hees and Ralf Rapp has predictions for
modifications to the $\Delta$(1232)$^{++}$ width at chemical and
kinetic freeze-out temperatures of T$_{\rm ch}$= 180 MeV T$_{\rm
kn}$= 120 MeV, which corresponds to densities of $\rho_{\rm ch}$ =
0.68 $\rho_{0}$ and $\rho_{\rm kn}$ = 0.12 $\rho_{0}$
\cite{hee04}. Calculations from M. Lutz \cite{lut01,lut02} predict
a mass shift and widths broadening of 40 MeV and 100 MeV for the
$\Sigma$(1385) and the $\Lambda$(1520) at a medium density of
$\rho$ = 1 $\rho_{0}$. Additional calculations which include the
density evolution of the fireball and further rescattering and
regeneration processes of the resonances need to be done to give
final answer on the measured mass position and width. The measured
data from SPS and RHIC do not show a mass shift or width
broadening in the errors of 5 MeV. At the moment we don't know if
this is a result that we would expect from theoretical
calculations.

\section*{References}

\bibliographystyle{amsunsrt}

\end{document}